\documentclass{sigplanconf}

\usepackage{amsmath}
\usepackage{cite}

  \usepackage[pdftex]{graphicx}
   \DeclareGraphicsExtensions{.pdf,.jpeg,.png}

\usepackage[tight,footnotesize]{subfigure}

\usepackage{url}

\begin{document}

\setlength{\pdfpageheight}{\paperheight}
\setlength{\pdfpagewidth}{\paperwidth}

\conferenceinfo{Promoto '13}{October 26, 2013, Indianapolis, Indiana, USA} 
\copyrightyear{2013} 
\copyrightdata{978-1-nnnn-nnnn-n/yy/mm} 
\doi{nnnnnnn.nnnnnnn}

% Uncomment one of the following two, if you are not going for the 
% traditional copyright transfer agreement.

%\exclusivelicense                % ACM gets exclusive license to publish, 
                                  % you retain copyright

%\permissiontopublish             % ACM gets nonexclusive license to publish
                                  % (paid open-access papers, 
                                  % short abstracts)

\titlebanner{banner above paper title}        % These are ignored unless
\preprintfooter{short description of paper}   % 'preprint' option specified.

\title{A Scratch-like visual programming system for Microsoft Windows Phone 8}
%\subtitle{Subtitle Text, if any}

\authorinfo{Annemarie Harzl, Philipp Neidh\"{o}fer,
Valentin Rock,
Maximilian Schafzahl, and 
Wolfgang Slany}
           {Graz University of Technology}
           {aharzl@ist.tugraz.at, philipp.neidhoefer@student.tugraz.at, valentin.rock@student.tugraz.at, mschafzahl@student.tugraz.at, wolfgang.slany@tugraz.at}
%\authorinfo{Name2\and Name3}
%           {Affiliation2/3}
 %          {Email2/3}

\maketitle

\section{Pocket Code}
Pocket Code is a free and open source mobile visual programming system for the Catrobat language\footnote{\url{http://developer.catrobat.org/}}. It allows users, starting from the age of eight, to develop games and animations with their smartphones. Children can create programs with their Android phone, iPhone, Windows Phone, or other smartphone with an HTML5 browser. No notebook or desktop computer is needed. Pocket Code is inspired by, but distinct from, the Scratch programming system developed by the Lifelong Kindergarten Group at the MIT Media Lab~\citep{Resnick2009}. Similar to Scratch, our aim is to enable children and teenagers to creatively develop and share their own software. The main differences between Pocket Code and Scratch are:
\begin{enumerate}
\item Support and integration of multi-touch mobile devices
\item Use of mobile device's special hardware (e.g., acceleration, compass, inclination)
\item No need for a traditional PC
\end{enumerate}
Currently there are more than 30 ongoing subprojects mostly aiming at extending Pocket Code’s functionality, e.g., a 2D physics engine that will make the programming of games similar to the popular Angry Birds type of games very easy, or an extension allowing to very easily record the screen as well as sound during execution of a program and to upload it to an online video sharing site. The high definition video will be created on our server and uploaded from there to avoid high costs and lengthy file transmissions for the children.

Pocket Code provides the functionality to share programs over all major platforms, i.e., Android, iOS, any HTML5 supported browser, and of course Windows Phone. The end user programmers, in our case mostly kids and teens, can learn from each other and share their ideas to develop new programs or games. To maximize the program and game pool, we developed a tool to convert already written Scratch programs to the Catrobat language. For the communication between our young developers and to provide them with updates and support, a forum was initiated. Within this forum every interested person can find consolidated knowledge and help, if help is needed.

\section{The Application}
\label{theApplication}

We are currently working on Integrated Development Environments (IDE) and interpreters for the Catrobat language on all major mobile platforms (Windows Phone, iOS, Android) and HTML5 browsers.
Pocket Code for Windows Platforms was initiated in 2011. The idea was and still is to provide Windows Phone users with the same functionality as users of Pocket Code on other mobile platforms have, but at the same time creating a unique user experience designed to fit in smoothly into the Windows Phone ecosystem. With the release of Windows 8 and its ability to provide a similar functionality as the Windows Phone platform does, the idea to cross-develop Pocket Code for Windows for Windows Phone, Windows RT, and upcoming platforms like Xbox One emerged. Since that moment our team has been working hard to write as much platform independent code as possible with the use of technologies like Model View ViewModel (MVVM\footnote{\url{http://www.galasoft.ch/mvvm}}) and Portable Class Libraries. 
In general, our Windows Phone solution is divided into two main parts: 
\begin{itemize}
	\item An IDE, which is written in C\# and strictly uses clean MVVM for an easy and least redundant development of views on multiple platforms.
	\item A Player, which is written in Visual C++ and uses Direct3D to ensure that Pocket Code projects get the best visual rendering performance on every device.	
\end{itemize}

Recently a new subproject was launched to build a paint application, called Pocket Paint, as Portable Class Library. Users of Pocket Paint can draw pictures, crop, flip, and fill images as they are accustomed to from desktop painting applications. Pocket Paint additionally supports transparent backgrounds important for allowing objects with arbitrary shapes.

\section{A Typical Usage Example}
\label{aTypicalUsageExample}

\begin{figure}[ht]
	\centering
	\includegraphics[width=3.3in]{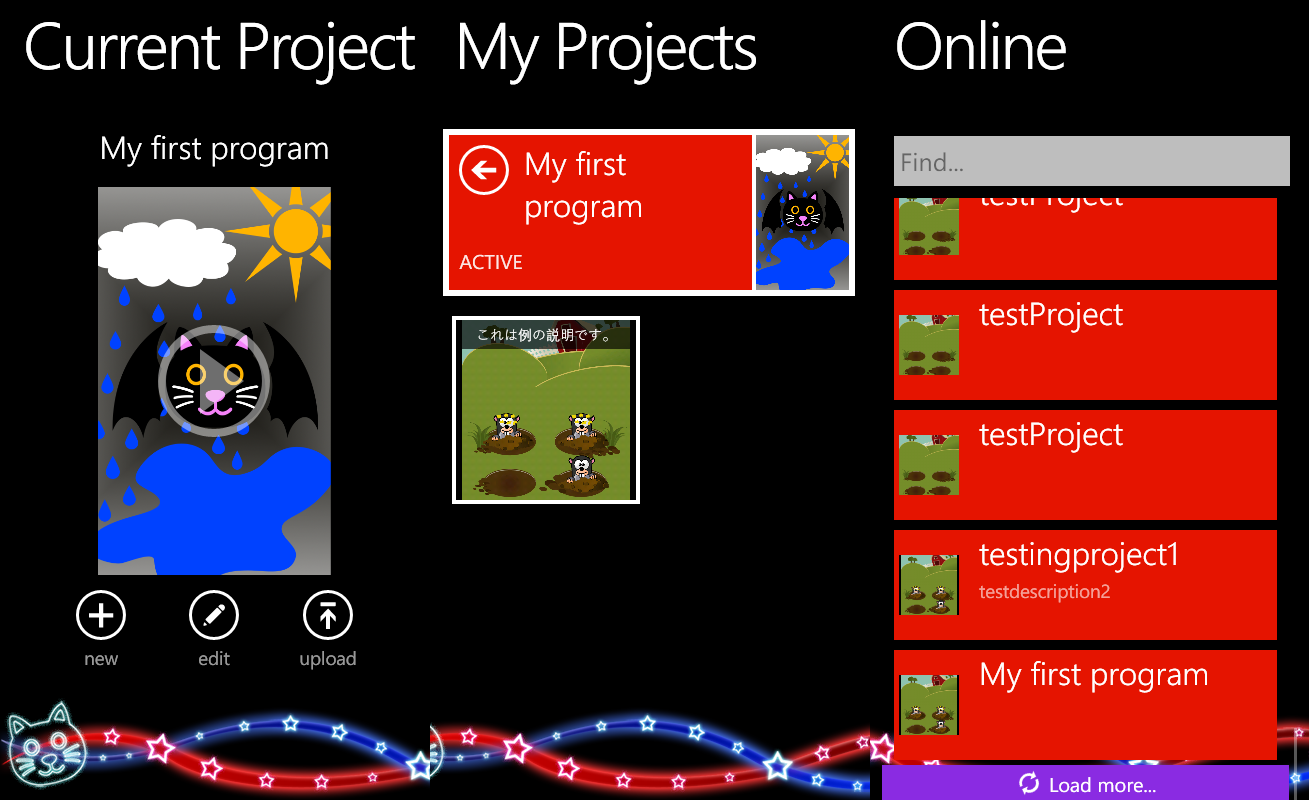}
	\caption{The main panorama view consisting of the current selected project, self created projects, and the online projects.}
	\label{fig:figure1}
\end{figure}

\begin{figure}[ht]
\centering
{
	\subfigure[Main page containing all objects of a project. Objects can be added, deleted and rearranged here.]
	{
		\includegraphics[width=1.48in]{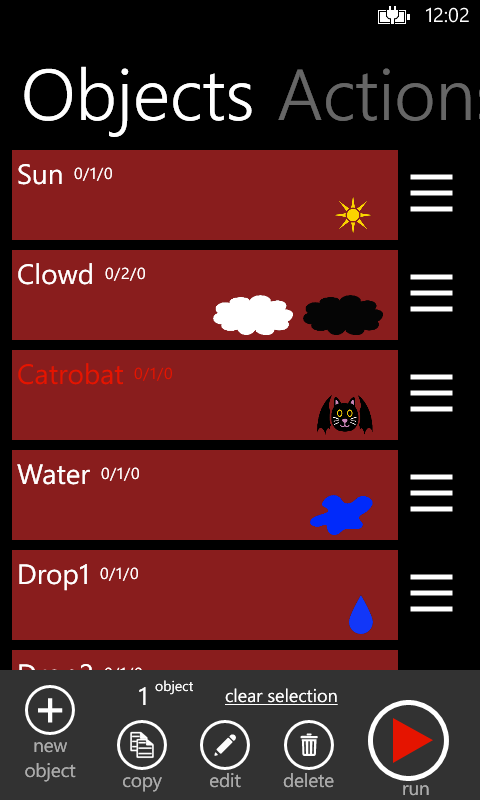}
		\label{fig:figure2}
		
	}
	\hfil
	\subfigure[Action bricks for an object.]
	{
		\includegraphics[width=1.48in]{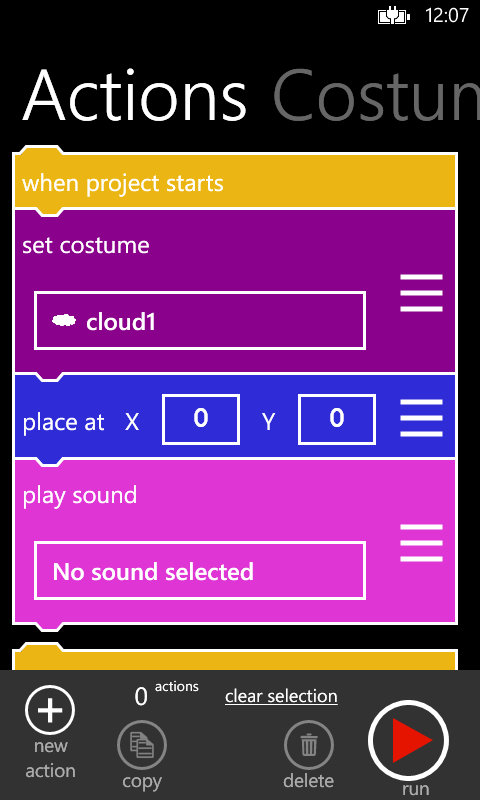}
		\label{fig:figure3}
	}
}
\caption{An object and a set of actions.}
\label{figure_objectsandactions}
\end{figure}

Users start with the main view of Pocket Code for Windows Phone (see Figure~\ref{fig:figure1}). There users can create new projects or open existing ones. A project can contain several objects (see Figure~\ref{fig:figure2}). Every object consists of three parts: costumes, sounds, and actions.

%Main view

\begin{figure}[ht]
	\centering
	{
	\subfigure[Some available bricks for the “motion” category.]
	{
		\includegraphics[width=1.48in]{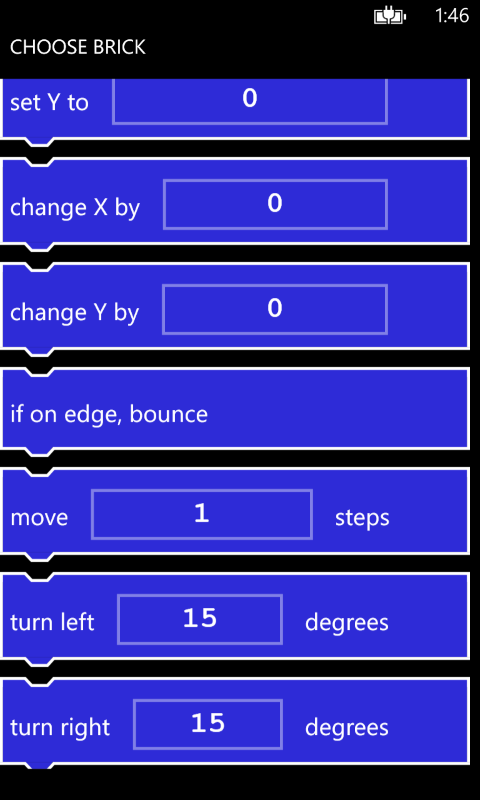}
		\label{fig:figure4}
	}
	\hfil
	\subfigure[For variable manipulation the Formula Editor is used. This figure shows the randomize function.]
	{
		\includegraphics[width=1.48in]{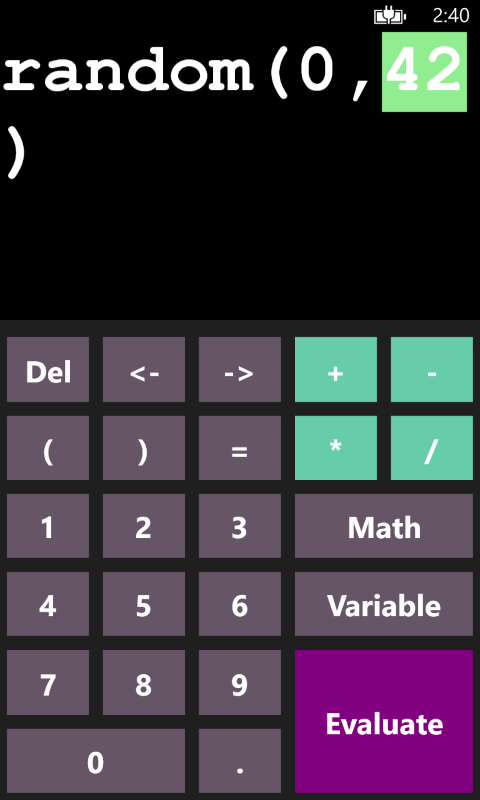}
		\label{fig:figure5}
	}

}
\caption{Brick list and Formula Editor.}
\label{figure_formulaandbricks}
\end{figure}

Costumes can basically be any kind of image, which can be captured with the internal camera, downloaded from the internet, or designed with Pocket Paint. 
Similar to costumes, sounds can be added to an object. Sounds can be retrieved from the internal storage, the internet, or they can be recorded with the internal microphone.
Theses costumes and sounds can then be used in the actions (see Figure~\ref{fig:figure3}). The set of all actions and their interactions represent a program. Actions are built with lego-like bricks, which have different functionalities and can be arranged and rearranged by dragging and dropping. Users are able to copy, edit, or delete these bricks. To interact with the program, simple OnTap events or sensor inputs from the device can be used. With the predefined lego-like bricks it is very easy to use logical statements, loops, and conditions, which are common in every programming language. This programming type using bricks avoids the problem of syntax errors. They simply are not possible, and thus possible user frustration is reduced. Modifications of the program can instantly be tested by executing the program. At present, end users have the possibility to choose between five categories of actions: motion, looks, sound, control, and variables. Each of these categories contains a set of predefined bricks (see Figure~\ref{fig:figure4}), which can be used to develop  programs.
To calculate variables the Formula Editor (see Figure~\ref{fig:figure5}) was developed. It has a pocket calculator like interface and gives users an overview over all available functions. Variables are available for all bricks in the same program and can be used for the calculation of the value of a parameter of a different brick.

%Objects and Actions

After creating a project, it can be uploaded to the Pocket Code community website and shared with friends and other users. Everyone can download these public projects, try, and use them, either starting a new project or editing projects. These modified and potentially enhanced projects can be shared on the same site in the spirit of open source software.

With our application children are able to animate pictures and create simple videos with these animations. Moreover they are able to build games with user interaction, tools such as a self-made compass, simulations, or any kind of other application.

Pocket Code is a fun and easy way to support the education of children concerning computers and to impart programming knowledge without bothering about the exact syntax of a programming languages. It helps children becoming creators of their own world instead of keeping them pure consumers of mobile technology.

%Formula Editor and Brick Types

\acks

Many thanks to the Catrobat team members\footnote{\url{http://catrob.at/credits}}.

% We recommend abbrvnat bibliography style.

\bibliographystyle{abbrvnat}

\bibliography{References}

% The bibliography should be embedded for final submission.

%\begin{thebibliography}{}
%\softraggedright

%\bibitem[Smith et~al.(2009)Smith, Jones]{smith02}
%P. Q. Smith, and X. Y. Jones. ...reference text...

%\end{thebibliography}

\end{document}